# Parameterization of the deuteron wave functions and form factors

**V. I. Zhaba**

Uzhgorod National University, Department of Theoretical Physics

54, Voloshyna St., Uzhgorod, UA-88000, Ukraine

E-mail address: viktorzh@meta.ua

**ABSTRACT**

Minimization of the number of numerically calculated coefficients for new analytical forms as a product of exponential function $r^1$ by the sum of the exponential terms $A_i*exp(-a_i*r^2)$ have been done. The optimum is $N$=10. Calculations have been done for realistic phenomenological Reid93 potential. Spherical $S_0$ and quadrupole $S_2$ form factors is parameterized by the coefficients for deuteron wave function in coordinate representation. The result $t_{20}(p)$ in wide area of momentas for Reid93 potential agreed well with the literature results for other potential nucleon-nucleon models, and with experimental data's of world collaborations and reviews. The obtained results will allow studying the deuteron electromagnetic structure, its form-factors, differential cross section of double scattering in more detail in future, and also for calculations the theoretical values of spin observables in *dp*-scattering.

*Keywords*: deuteron, wave function, form factors, tensor polarization, parametrization.

## 1. INTRODUCTION

Deuteron is the most elementary nucleus. He consists of the two strongly interacting elementary particles: a proton and a neutron. The simplicity and evidentness of the deuteron's structure makes it a convenient laboratory for studying and modeling nucleon-nucleon forces. Now, deuteron has been well investigated both experimentally and theoretically.

It should also be noted that such potentials of the nucleon-nucleon interaction as Bonn, Moscow, Nijmegen group potentials (NijmI, NijmII, Nijm93 [1]), Argonne v18, Paris, NLO, NNLO and N3LO, Idaho N3LO or Oxford have quite a complicated structure and cumbersome representation. Example, the original potential Reid68 was parameterized on the basis of the phase analysis by Nijmegen group and was called as updated regularized version - Reid93. The parameterization was done for 50 parameters of the potential, where value $\chi^2/N_{data}$=1.03 [2].

Besides, the deuteron wave function (DWF) in coordinate space can be presented as a table: through respective arrays of values of radial wave functions. It is sometimes quite difficult and inconvenient to operate with such arrays of numbers during numerical calculations. And the program code for numerical calculations is bulky, overloaded and unreadable. Therefore, it is feasible to obtain simpler and comfortabler analytical forms of DWF representation. It is further possible on the basis to calculate the form factors and tensor polarization, characterizing the deuteron structure.

DWFs in a convenient form are necessary for use in calculations of polarization characteristics of the deuteron, as well as to evaluate the theoretical values of spin observables in *dp* scattering.

The main objectives of the research in this paper are to parameterize, calculate and analyze the deuteron form factors, what obtained by coefficients for new analytical forms DWF in coordinate representation.

## 2. ANALYTICAL FORM OF THE DEUTERON WAVE FUNCTION

In 2000-x years are new analytical forms of deuteron wave function. Except the mentioned parameterization, in literature there is one more analytical form [3] for DWFs

$$\begin{cases} u(r) = r \sum_{i=1}^{N} A_i e^{-a_i r^2}, \\ w(r) = r \sum_{i=1}^{N} B_i e^{-b_i r^2}. \end{cases} \quad (1)$$

The accuracy of parameterization (1) is characterized by:

$$\chi^2 = \frac{1}{n-p} \sum_{i=1}^{N} \left( y_i - f(x_i; a_1, a_2, ..., a_p) \right)^2, \quad (2)$$

where $n$ - the number of points of the array $y_i$ of the numerical values of DWF in the coordinate space; $f$ - approximating function of $u$ (or $w$); $a_1, a_2, ..., a_p$ - parameters; $p$ - the number of parameters (coefficients in the sums of formulas (1)). Hence, $\chi^2$ is determined not only by the shape of the approximating function $f$, but also by the number of the selected parameters.

If we consider normalization $\int (u^2 + w^2) dr = 1$ for DWFs (1), we can write this condition using the corresponding coefficients as

$$\frac{\sqrt{\pi}}{2^{7/2}} \sum_{i=1}^{N} \left( \frac{A_i^2}{a_i^{3/2}} + \frac{B_i^2}{b_i^{3/2}} \right) = 1.$$

Based on the known DWFs one can calculate the deuteron properties:
the deuteron radius $r_m$

$$r_d = \frac{1}{2} \left\{ \int_0^\infty r^2 \left[ u^2(r) + w^2(r) \right] dr \right\}^{1/2};$$

the quadrupole moment $Q_d$

$$Q_d = \frac{1}{20} \int_0^\infty r^2 w(r) \left[ \sqrt{8} u(r) - w(r) \right] dr;$$

the magnetic moment $\mu_d$

$$\mu_d = \mu_s - \frac{3}{2} (\mu_s - \frac{1}{2}) P_D;$$

the D- state probability $P_D$

$$P_D = \int_0^\infty w^2(r) dr;$$

the "D/S- state ratio" $\eta$

$$\eta = A_D / A_S.$$

In a formula for $\mu_d$ size $\mu_s = \mu_n + \mu_p$ is the sum of the magnetic moments of a neutron and proton. Value of the calculated magnetic moment of a deuteron is given in nuclear magnetons $\mu_N$.

In Table 1 these values are indicated depending on the number $N$. Dependence of $\chi^2$ on the number $N$ of expansion terms for parameterization (1) is shown separately for the functions $u(r)$ and $w(r)$. The optimum is $N=10$. Based on the known DWFs (1) and them coefficients one can calculate the deuteron properties: the D- state probability $P_D$, deuteron radius $r_m$, the quadrupole moment $Q_d$, the magnetic moment $\mu_d$ and the "D/S- state ratio" $\eta$ They are in good agreement with the theoretical [1] and experimental [4] data's. The coefficients for DWFs in coordinate space for Reid93 potential have been numerically calculated (Table 2).

**Table 1.** $\chi^2$ for DWFs (1) and deuteron properties.

| $N$ | $\chi^2$ for $u(r)$ | $\chi^2$ for $w(r)$ | $P_D$ (%) | $r_m$ (fm) | $Q_d$ (fm$^2$) | $\mu_d$ | $\eta$ |
|---|---|---|---|---|---|---|---|
| 1 | 3.98E-03 | 2.11E-04 | 5.356 | 1.4720 | 0.1658 | 0.84929 | 0.00010 |
| 2 | 8.26E-04 | 1.47E-04 | 5.510 | 1.9009 | 0.1505 | 0.84841 | 0.00050 |
| 3 | 7.60E-04 | 1.19E-04 | 5.467 | 1.9678 | 0.2573 | 0.84866 | 0.00734 |
| 4 | 7.62E-04 | 1.20E-04 | 5.467 | 1.9678 | 0.2573 | 0.84866 | 0.00734 |
| 5 | 7.59E-04 | 1.20E-04 | 5.467 | 1.9705 | 0.2575 | 0.84866 | 0.00690 |
| 6 | 7.57E-04 | 6.16E-06 | 5.697 | 1.9673 | 0.2413 | 0.84734 | 0.00137 |
| 7 | 7.58E-04 | 3.15E-07 | 5.696 | 1.9679 | 0.2632 | 0.84735 | 0.01811 |
| 8 | 7.66E-04 | 4.18E-08 | 5.701 | 1.9681 | 0.2691 | 0.84732 | 0.02790 |
| 9 | 9.63E-08 | 4.14E-08 | 5.700 | 1.9687 | 0.2694 | 0.84733 | 0.02731 |
| **10** | **1.93E-08** | **3.60E-09** | 5.699 | 1.9667 | 0.2700 | 0.84734 | 0.02548 |
| 11 | 9.74E-08 | 3.86E-09 | 5.700 | 1.9682 | 0.2702 | 0.84733 | 0.02534 |
| 12 | 5.76E-08 | 5.96E-09 | 5.700 | 1.9683 | 0.2702 | 0.84733 | 0.02584 |
| 13 | 4.65E-08 | 6.79E-09 | 5.698 | 1.9679 | 0.2699 | 0.84734 | 0.02602 |
| 14 | 9.09E-08 | 2.00E-08 | 5.698 | 1.9681 | 0.2697 | 0.84734 | 0.02703 |
| 15 | 8.46E-08 | 2.28E-08 | 5.698 | 1.9677 | 0.2698 | 0.84734 | 0.02727 |
| Theor. | - | - | 5.699 | 1.969 | 0.2703 | - | 0.0251 |
| Exp. | - | - | - | 1.9753 | 0.2859 | 0.85744 | 0.0256 |

**Table 2.** Coefficients $A_i$, $a_i$, $B_i$, $b_i$.

| $i$ | $A_i$ | $a_i$ | $B_i$ | $b_i$ |
|---|---|---|---|---|
| 1 | -4.86170097 | 3.36265132 | -0.23984123 | 2.85287828 |
| 2 | 0.00843757 | 0.00738593 | 0.00137949 | 0.01153549 |
| 3 | 2.16784051 | 3.46888099 | 0.11562677 | 0.60245431 |
| 4 | 0.27790561 | 0.12953766 | 0.27570090 | 0.05939491 |
| 5 | -0.34177140 | 0.07689243 | -0.57255823 | 0.09054895 |
| 6 | 0.26478838 | 0.02964324 | 0.30262540 | 0.08431200 |
| 7 | 2.16784051 | 3.46888099 | 0.09071340 | 0.23131272 |
| 8 | 0.27170156 | 0.44391265 | 0.30310748 | 0.08431420 |
| 9 | -0.34573458 | 0.03533723 | -0.57983043 | 0.06756505 |

| 10 | 0.44597882 | 0.05815706 | 0.30327344 | 0.08431495 |

The obtained deuteron wave functions (Fig. 1 and 2) do not contain any superfluous knots for *N*=10. DWFs for the «worst» approximations with *N*=2; 5 are also specified.

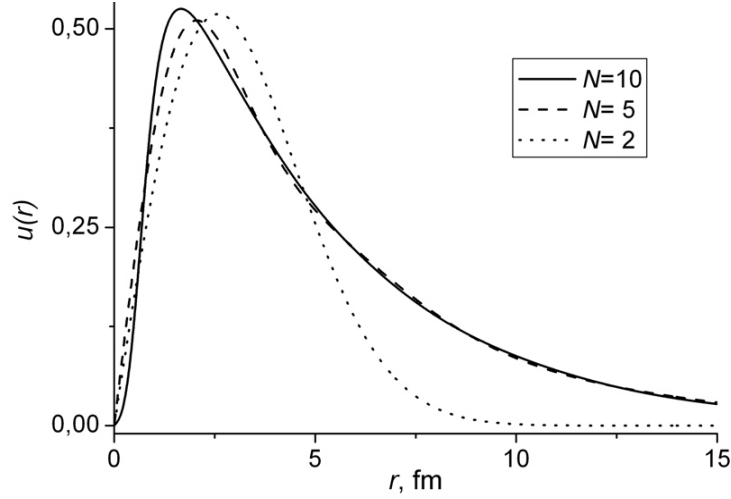

**Figure 1.** Deuteron wave function *u(r)*.

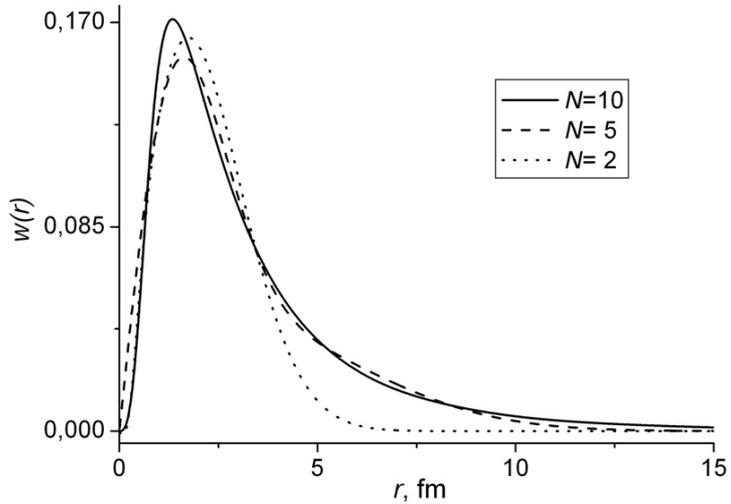

**Figure 2.** Deuteron wave function *w(r)*.

## 3. DEUTERON FORM FACTORS AND TENSOR POLARIZATION

The polarization characteristics of deuteron can be calculated based on the received DWFs (1). These include tensor polarization, polarization transmission, tensor analyzing power et al. Tensor polarization $t_{20}$ [4, 5] of the repulsed deuterons is determined using the form factors $G_i$:

$$t_{20}(p) = -\frac{1}{\sqrt{2}S}\left(\frac{8}{3}\eta G_C(p)G_Q(p) + \frac{8}{9}\eta G_Q^2(p) + \frac{1}{3}\eta\left[1 + 2(1+\eta)tg^2\left(\frac{\theta}{2}\right)\right]G_M^2(p)\right), \quad (3)$$

where $\eta = \frac{p^2}{4m_d^2}$; $m_d$=1875.63 MeV - deuteron mass; $S = A(p) + B(p)tg^2\left(\frac{\theta}{2}\right)$. Here *A(p)* and *B(p)* - functions of the electric and magnetic structure:

$$A = G_C^2 + \frac{8}{9}\eta^2 G_Q^2 + \frac{2}{3}\eta G_M^2; \qquad B = \frac{4}{3}\eta(1+\eta)G_M^2.$$

Charge $G_C(p)$, quadrupole $G_Q(p)$ and magnetic $G_M(p)$ deuteron form factors contain information about the electromagnetic properties of the deuteron [5]:

$$G_C = G_{EN} D_C; \qquad G_Q = G_{EN} D_Q; \qquad G_M = \frac{m_d}{2m_p}(G_{MN} D_M + G_{EN} D_E) \qquad (4)$$

Here $G_{Ep}$, $G_{En}$ ($G_{Mp}$, $G_{Mn}$) - neutron and proton electric (magnetic) form factors.

It is convenient to divide the spherical $S_0$ and quadrupole $S_2$ form factors [6] into two parts which correspond to different multiplicities of the deuteron wave function for S- and D-states:

$$S_0(p/2) = S_0^{(1)} + S_0^{(2)}; \qquad (5)$$

$$S_2(p/2) = 2S_2^{(1)} - \frac{1}{\sqrt{2}} S_2^{(2)}; \qquad (6)$$

where elementary spherical and quadrupole form factors

$$S_0^{(1)} = \int_0^\infty u^2 j_0 dr; \qquad S_0^{(2)} = \int_0^\infty w^2 j_0 dr; \qquad (7)$$

$$S_2^{(1)} = \int_0^\infty uw j_2 dr; \qquad S_2^{(2)} = \int_0^\infty w^2 j_2 dr. \qquad (8)$$

In formulas (7) and (8) $j_0$, $j_2$ - the spherical Bessel functions from the argument $pr/2$. In [7] formulas (3) and (4) have been written down with argument $pr$.

Then components of deuteron form factors $G_C$, $G_Q$, $G_M$ in definitions elementary spherical and quadrupole form factors (7) and (8) will get the form as

$$D_C = S_0^{(1)} + S_0^{(2)}; \qquad D_Q = \frac{3}{\sqrt{2}\eta}\left(S_2^{(1)} - \frac{1}{\sqrt{8}} S_2^{(2)}\right);$$

$$D_M = 2\left(S_0^{(1)} - \frac{1}{2} S_0^{(2)} + \frac{1}{\sqrt{2}} S_2^{(1)} + \frac{1}{2} S_2^{(2)}\right); \qquad D_E = \frac{3}{2}\left(S_0^{(2)} + S_2^{(2)}\right).$$

Then formulas (7) and (8) for DWFs (1) will be written down in parameterization forms

$$S_0^{(1)} = \frac{\sqrt{\pi}}{4} \sum_{i,j=1}^N \frac{A_i A_j e^{-\frac{p^2}{16(a_i+a_j)}}}{(a_i+a_j)^{3/2}}; \qquad S_0^{(2)} = \frac{\sqrt{\pi}}{4} \sum_{i,j=1}^N \frac{B_i B_j e^{-\frac{p^2}{16(b_i+b_j)}}}{(b_i+b_j)^{3/2}}; \qquad (9)$$

$$S_2^{(1)} = -2\sqrt{\pi} \sum_{i=1}^N A_i B_i \left(\frac{(p^2 + 24(a_i+b_i))e^{-\frac{p^2}{16(a_i+b_i)}}}{8p^2(a_i+b_i)^{3/2}} + \frac{6\sqrt{\pi}}{p^3} erf\left(\frac{p}{4\sqrt{a_i+b_i}}\right)\right); \qquad (10)$$

$$S_2^{(2)} = -2\sqrt{\pi} \sum_{i,j=1}^N B_i B_j \left(\frac{(p^2 + 24(b_i+b_j))e^{-\frac{p^2}{16(b_i+b_j)}}}{8p^2(b_i+b_j)^{3/2}} + \frac{6\sqrt{\pi}}{p^3} erf\left(\frac{p}{4\sqrt{b_i+b_j}}\right)\right). \qquad (11)$$

where $erf(z) = \frac{2}{\sqrt{\pi}} \int_0^z e^{-t^2} dt$ - error function.

So, a form factors by formulas (9)-(11) is determined by the coefficients for decomposition of DWFs in coordinate representation. Spherical and quadrupole form factors

$S_i^{(j)}$ are shown in Fig. 3. Calculations are made for the original dipole fit (DFF) for the proton and neutron form factors [8]:

$$G_{Ep} = F_N; \qquad G_{En} = 0; \qquad G_{Mp} = \mu_p G_{Ep}; \qquad G_{Mn} = \mu_n G_{Ep};$$

where $F_N(p^2) = \left(1 + \dfrac{p^2}{18.235\, fm^{-2}}\right)^{-2}$ - nucleon form factor. In Fig. 3 is specified for form factors near the beginning of coordinate the behaviour in logarithmic and usual scale. Zeros of form factors are distinctly visible.

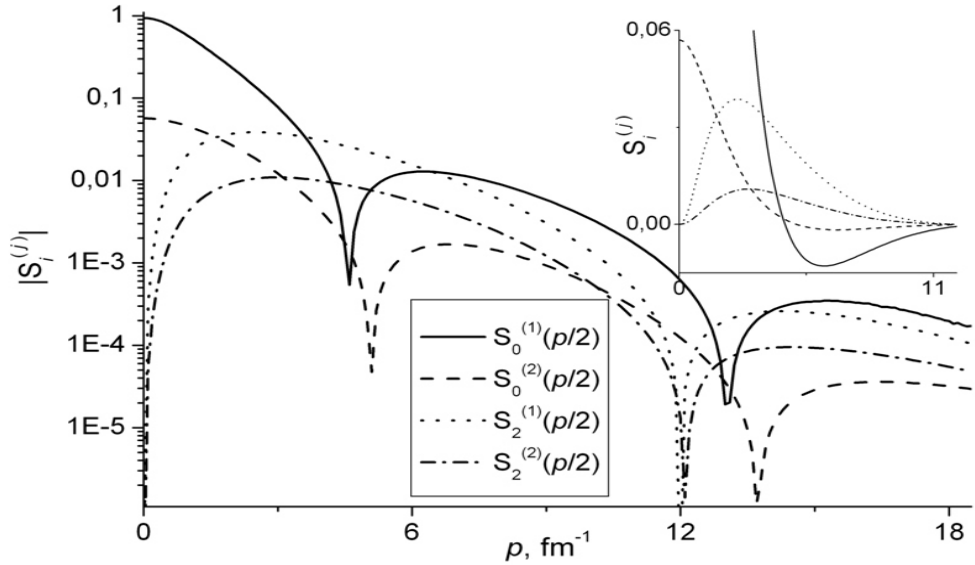

**Figure 3.** Spherical and quadrupole form factors $S_i^{(j)}$ for Reid93 potential.

The deuteron tensor polarization $t_{20}$ (Fig. 4) has been calculated based on the received DWFs (1) and form factors (9)-(11). A detailed comparison of the obtained values of $t_{20}$ (the scattering angle $\theta=70^0$) for Reid93 potential with the up-to-date experimental data's of Bates [9, 10], BLAST [11, 12], JLab [13], NIKHEF [14, 15], VEPP-3 [16-18], Saclay [19] collaborations and Abbott [20], Boden [21], Garcon [10] reviews. There is a good agreement is for the momentas $p$=1-4 fm$^{-1}$.

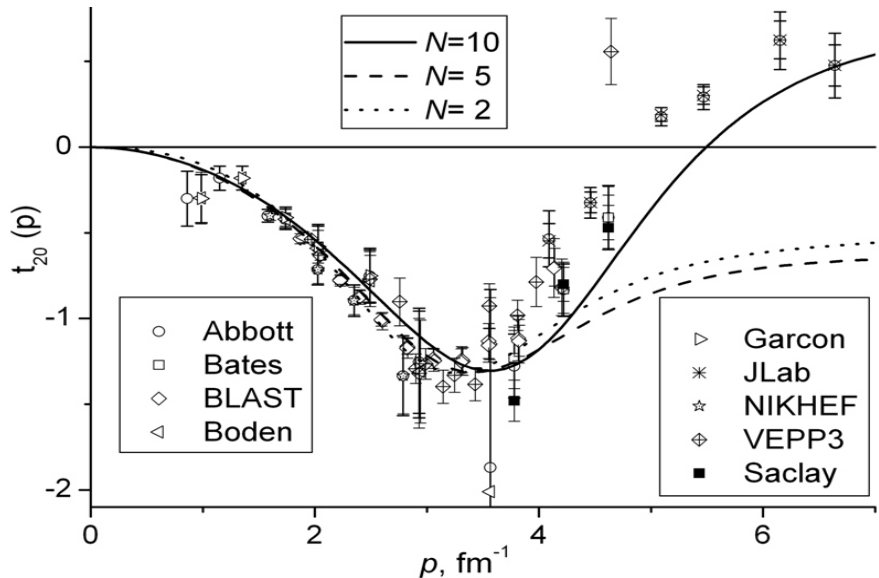

**Figure 4.** Tensor polarization $t_{20}$.

The calculated value $t_{20}(p)$ is in good agreement with the results of works, where the theoretical calculations have been conducted: with available data's [10] for the Paris, Argonne v14 and Bonn-E potentials and with data [22] for Moscow, NijmI, NijmII, CD-Bonn and Paris potentials. It is in a good agreement with the value $t_{20}(p)$ calculated in [12] for elastic *ed*-scattering for models with the inclusion of nucleon isobaric component, within light-front dynamics and quark cluster model, for Bonn-A, Bonn-B and Bonn-C, Bonn Q, Reid-SC and Paris A-VIS potentials. Besides, $t_{20}(p)$ coincide well with the results according to the effective field theory [12].

## 4. CONCLUSIONS

1. Parameterization for DWFs in coordinate representation of the form (1) has been used and the number of expansion coefficients has been minimized. Dependence of $\chi^2$ on the number *N* of expansion terms parameterization (1) is shown separately for the functions *u(r)* and *w(r)*. The optimum is *N*=10. Minimization of the number of numerically calculated coefficients for new analytical forms (1) have been done. Static properties of the deuteron ($P_D$, $r_m$, $Q_d$, $\mu_d$, $\eta$), obtained by DWFs for Reid93 potential, have been numerically calculated. The resulting wave functions do not contain any extra knots.

2. Spherical and quadrupole form factors is determined by the coefficients for decomposition of DWFs (1) in coordinate representation. These will be written down in parameterization forms (9)-(11).

3. The result $t_{20}(p)$ in wide area of momentas for Reid93 potential agreed well with the literature results for other potential nucleon-nucleon models, and with experimental data of world collaborations (Bates, BLAST, JLab, NIKHEF, VEPP-3, Saclay) and reviews (Abbott, Boden, Garcon).

4. The resulting DWFs for the group of potential models can be applied to calculate polarization characteristics of the deuteron (sensitivity tensor component to polarization of deuterons $T_{20}$, polarization transmission $K_0$ etc.) and compare them with theoretical calculations, as well as with the experimental data [23]. The results will allow studying the deuteron electromagnetic structure and differential cross section of double scattering in more detail in future. In a convenient form analytical forms for spherical $S_0$ and quadrupole $S_2$ deuteron form factors are necessary for use in calculations of the theoretical values of tensor analyzing power $A_{yy}$, vector analyzing power $A_y$, vector-vector and the tensor-tensor polarization transfer and spin-flip cross sections for the reaction $^{12}C(d, d)X$ [6].